\PassOptionsToPackage{pdfpagelabels=false}{hyperref}
\documentclass[fleqn,usenatbib,usedcolumn]{mnras}
\usepackage[british]{babel}             

\usepackage{newtxmath,newtxtext}

\usepackage[T1]{fontenc}

\usepackage{graphicx}    
\usepackage{amsmath}    
\usepackage{amssymb}    
\usepackage{pdflscape}
\usepackage{etoolbox}
\usepackage{CJK}
\makeatletter
\patchcmd\@combinedblfloats{\box\@outputbox}{\unvbox\@outputbox}{}{%
   \errmessage{\noexpand\@combinedblfloats could not be patched}%
}%
 \makeatother
   
\newcommand{\feh}{\ensuremath{[\textrm{Fe}/\textrm{H}]}}
\newcommand{\bafe}{\ensuremath{\textrm{Ba}^\star}}
\newcommand{\mgfe}{\ensuremath{\textrm{Mg}^\star}}
\newcommand{\lireal}{\ensuremath{[\textrm{Li}/\textrm{Fe}]}}
\newcommand{\bareal}{\ensuremath{[\textrm{Ba}/\textrm{Fe}]}}
\newcommand{\life}{\ensuremath{\textrm{Li}^\star}}
\newcommand{\logg}{\ensuremath{\log g}}
\newcommand{\vsini}{\ensuremath{v\sin i}}
\newcommand{\teff}{\ensuremath{T_\textrm{eff}}}
\newcommand{\loggdag}{\ensuremath{\log g^\star}}
\newcommand{\vsinidag}{\ensuremath{v\sin i^\star}}
\newcommand{\teffdag}{\ensuremath{T_\textrm{eff}^\star}}
\newcommand{\fehdag}{\ensuremath{[\textrm{Fe}/\textrm{H}]^\star}}

\begin{document}
\begin{CJK*}{UTF8}{gbsn}
\label{firstpage}
\pagerange{\pageref{firstpage}--\pageref{lastpage}}


\title[GALAH and the Clouds]{The GALAH and \textit{TESS}-HERMES surveys: high-resolution spectroscopy of luminous supergiants in the Magellanic Clouds and Bridge}

\author[J. D. Simpson et al.]{Jeffrey D. Simpson$^{1}$\thanks{Email: \texttt{jeffrey.simpson@aao.gov.au}},
Dennis~Stello$^{2,3,4}$,
Sanjib~Sharma$^{5}$,
Yuan-Sen~Ting$^{6,7,8}$,\newauthor
David~M.~Nataf$^{9}$,
Gary~Da~Costa$^{10}$,
Robert A.Wittenmyer$^{11}$,
Jonathan~Horner$^{11}$,\newauthor
Sarah~L.~Martell$^{3}$,
Geraint~F.~Lewis$^{5}$,
Gayandhi~M.~De~Silva$^{1,5}$,
Peter~L.~Cottrell$^{12,13}$,\newauthor
Martin~Asplund$^{2,10}$,
Joss~Bland-Hawthorn$^{2,5}$,
Sven Buder$^{14}$,
Valentina~{D'Orazi}$^{15}$,\newauthor
Ly~Duong$^{2,10}$,
Ken~C.~Freeman$^{10}$,
Janez~Kos$^{5}$,
Jane~Lin$^{2,10}$,
Karin~Lind$^{14,16}$,\newauthor
Katharine.~J.~Schlesinger$^{10}$,
Daniel~B.~Zucker$^{17}$,
Toma\v{z}~Zwitter$^{18}$,
Prajwal~R.~Kafle$^{19}$,\newauthor
Shourya Khanna$^{2,10}$,
Thomas~Nordlander$^{2,10}$
\\
$^{1}$Australian Astronomical Observatory, 105 Delhi Rd, North Ryde, NSW 2113, Australia\\
$^{2}$Centre of Excellence for Astrophysics in Three Dimensions (ASTRO-3D), Australia\\
$^{3}$School of Physics, UNSW, Sydney, NSW 2052, Australia\\
$^{4}$Stellar Astrophysics Centre, Department of Physics and Astronomy, Aarhus University, DK-8000, Aarhus C, Denmark\\
$^{5}$School of Physics, UNSW, Sydney, NSW 2052, Australia\\
$^{6}$Institute for Advanced Study, Princeton, NJ 08540, USA \\
$^{7}$Department of Astrophysical Sciences, Princeton University, Princeton, NJ 08544, USA \\
$^{8}$Observatories of the Carnegie Institution of Washington, 813 Santa Barbara Street, Pasadena, CA 91101, USA \\
$^{9}$Center for Astrophysical Sciences and Department of Physics and Astronomy, The Johns Hopkins University, Baltimore, MD 21218, USA\\
$^{10}$Research School of Astronomy \& Astrophysics, Australian National University, ACT 2611, Australia\\
$^{11}$University of Southern Queensland, Toowoomba, Queensland 4350, Australia\\
$^{12}$School of Physical and Chemical Sciences, University of Canterbury, New Zealand\\
$^{13}$Monash Centre for Astrophysics, School of Physics and Astronomy, Monash University, Australia\\
$^{14}$Max Planck Institute  for Astronomy (MPIA), Koenigstuhl 17, 69117 Heidelberg, Germany\\
$^{15}$Istituto Nazionale di Astrofisica, Osservatorio Astronomico di Padova, vicolo dell'Osservatorio 5, 35122, Padova, Italy \\
$^{16}$Department of Physics and Astronomy, Uppsala University, Box 517, SE-751 20 Uppsala, Sweden\\
$^{17}$Department of Physics and Astronomy, Macquarie University, Sydney, NSW 2109, Australia \\
$^{18}$Faculty of Mathematics and Physics, University of Ljubljana, Jadranska 19, 1000 Ljubljana, Slovenia\\
$^{19}$ICRAR, The University of Western Australia, 35 Stirling Highway, Crawley, WA 6009, Australia \\
}

\date{Accepted XXX. Received YYY; in original form ZZZ}

\pubyear{2018}

\maketitle
\end{CJK*}

\begin{abstract}
We report the serendipitous observations of 571 luminous supergiants in the Magellanic Clouds by the spectroscopic GALAH and \textit{TESS}-HERMES surveys: 434 stars in the Large Magellanic Cloud and 137 in the Small Magellanic Cloud. We also find one star that appears associated with structured star formation in the Magellanic Bridge. Both of these surveys are aimed at the local volume of the Galaxy but have simple, magnitude-limited selection functions that mean they include some observations of luminous extra-Galactic stars. The surveys determine stellar parameter and abundances using \textit{The Cannon}, a data-driven generative modelling approach. In this work, we explore the results from \textit{The Cannon} when it is fed the spectra of these intrinsically luminous supergiants in the Magellanic Clouds, which are well outside the normal bounds of \textit{The Cannon}'s training set. We find that, although the parameters are astrophysically incorrect, the \vsini\ and the abundances of lithium, barium, and magnesium are excellent discriminants of these stars. It shows that in the future, with an expanded training set, it should be possible to determine accurate values for these types of stars.
\end{abstract}




\section{Introduction}
\label{sec:intro}
The Large and Small Magellanic Clouds (LMC and SMC) are the two most massive satellite galaxies of the Milky Way, and provide a unique laboratory for studying a wide range of astrophysical processes. Unlike most other external galaxies, individual stars can be resolved and studied with moderate-sized telescopes. As well as the Magellanic Clouds (MC) themselves, there are several related structures including the Magellanic Bridge \citep[discovered by][]{Hindman1963}, the leading stream of neutral hydrogen known as the leading arms \citep{Putman1998} and the trailing stream of gas, known as the Magellanic Stream \citep[discovered by][]{Mathewson1974}. For a comprehensive review see \citet{DOnghia2015}. This provides astronomers with the closest example of interacting galaxies, and the MC have been the target of numerous dedicated studies at multiple wavelengths. The brightest MC stars have the same apparent magnitudes as stars studied by large-scale Galactic stellar surveys, so it is possible these extra-Galactic stars will be targeted deliberately as part of such spectroscopic surveys e.g., RAVE \citep{Munari2009}, APOGEE-2 \citep{Zasowski2017}, or appear serendipitously as in GALAH (this work) and \textit{TESS}-HERMES \citep{Sharma2017a}.

In this work, we identify MC stars that were serendipitously observed by two surveys using the High Efficiency and Resolution Multi-Element Spectrograph (HERMES) \citep{Sheinis2015} on the Anglo-Australian Telescope (AAT): the GALactic Archaeology with HERMES (GALAH) survey and the \textit{TESS}-HERMES survey. The GALAH survey is a high-resolution stellar spectroscopic study of the local volume of the Galaxy, with the primary aim of undertaking an ambitious chemical tagging project \citep{DeSilva2015a,Martell2017a}. The \textit{TESS}-HERMES survey \citep{Sharma2017a} aims to determine spectroscopic parameters for stars in the \textit{TESS} mission's Southern Continuous Viewing Zone \citep{Ricker2014}. For both surveys, the selection functions are simple, with no colour cuts, and are just magnitude limited, with the faintest stars observed have $V\approx14$. Although there are $\sim40$ dwarf galaxies within the sky coverage of GALAH and \textit{TESS}-HERMES, it is reasonable to assume that the only extra-galactic stars that will be observed are from the MC due to the target density and the number of observations made by the surveys in and around the MC.

The input catalogue for the GALAH and \textit{TESS}-HERMES surveys use a $V$ magnitude estimated from 2MASS $J,K_S$ magnitudes
\begin{equation}\label{eq:mag}
	V_{J,K} = K_S + 2.0(J-K_S+0.14)+0.382e^{2(J-K_S-0.2)}.
\end{equation}
This estimate was calibrated for the predominant stars in the local Galactic volume probed by GALAH: dwarfs, turnoff stars and first-ascent RGB stars with modest reddening. Using MIST isochrones \citep{Dotter2016,Choi2016,Paxton2011,Paxton2013,Paxton2015} for RGB, HB and AGB stars with MC-like properties (aged 10--100~Myr with $\feh=-0.3$ and $A_V=0.30$), we estimate that the $V-V_{J,K}\approx0.5$, hence, the faintest stars observed of these evolutionary stages are actually $V\approx13.5$. 

With distance moduli of $m-M=18.49$ for the LMC \citep{Pietrzynski2013} and $m-M=18.95$ for the SMC \citep{Graczyk2013}, only stars with absolute magnitudes $M_V<-5$ and $M_V<-5.5$ respectively can be observed in the MC. This limits us to the brightest supergiants: Wolf-Rayet stars, hot OBA-type stars, and cool supergiants. The spectra of these types of stars are not well-suited to the common parameter estimation pipeline used by both surveys, which is tuned to determine elemental abundances for stars with $4000\mathrm{~K}<\teff<8000$~K and $0<\logg<6$ \citep[for a comprehensive discussion of the parameter determination, see][]{Buder2018}. The pipeline uses a two-step process to determine stellar parameters and elemental abundances --- henceforth collectively known as `labels' --- for the stars observed. In the first step, a training set of $\sim$10000 stars is selected that have high signal-to-noise and cover the expected parameter space. These spectra are analyzed with Spectroscopy Made Easy \citep[SME;][]{Valenti1996,Piskunov2017} to determine their labels using classical spectrum synthesis methods. In the second step, these training set spectra and stellar parameters are fed into \textit{The Cannon} \citep{Ness2015}, a data-driven generative modelling approach to label determination. \textit{The Cannon} builds a quadratic model at each pixel (ie., wavelength step) of the normalised spectrum as a function of the labels. This model is then used to determine the labels for the bulk of the spectra in a computationally short amount of time.

The primary aim of \textit{The Cannon} is to produce labels that are both \textit{precise} and \textit{accurate}. With no colour cuts to avoid the hottest and coolest stars in these surveys, there will be spectra acquired for which the true parameters are well outside of the training set's bounds, and therefore \textit{The Cannon} will not be able to produce accurate labels. As such, there is robust flagging of the reliability of the labels to inform the end user whether these labels should be used for detailed abundance studies. But it is not unreasonable to expect that there should be coherence of unreliable labels for a given group of stars, such as O supergiants in the MC. It is this fact that we wish to exploit in this work to identify stars that belong to the MC, and the surrounding structures like the Magellanic Bridge.

The Magellanic Bridge was first discovered as a stream of HI gas connecting the two MC \citep{Hindman1963}. Later work identified that it had a stellar counterpart \citep[e.g.,][]{Irwin1990}, with the ages of these stars suggesting that they must have formed in situ, rather than being stripped from the MC. Recent work from \citet{Carrera2017} has, for the first time, found older red giant stars that they claim were tidally stripped from the LMC by the SMC. \citet{Belokurov2017} found stellar tidal tails around the LMC and the SMC using RR Lyrae stars, which supports the model of interaction between the MC and that the old stars are not in the same location as the young stars in the Magellanic Bridge. The ongoing SMASH survey is using DECAM to map this region as well \citep{Nidever2017}. It would be of great interest if Magellanic Bridge stars were serendipitously observed by GALAH or \textit{TESS}-HERMES.

Data-driven classification methods have already been applied to GALAH and \textit{TESS}-HERMES spectra: \citet{Traven2017} used the recently developed dimensionality reduction technique t-SNE (t-distributed Stochastic Neighbour Embedding) to represent complex spectral morphology on a two-dimensional map, and to classify stars with ``unusual'' spectra, and to enable the flagging of potentially problematic spectra. Searching for outliers with machine learning techniques has also been applied to other large surveys \citep[e.g.,][]{Reis2017}. Here, we use the labels, not the pixel information of the spectrum, for the classification. 

In this work we discuss the observations and their analysis (Section \ref{sec:data_reduction}); an initial photometric criterion (Section \ref{sec:skymapper_select}); the final criteria using kinematics and \textit{Cannon} labels (Section \ref{sec:final_criteria}); the reliability of the method (Section \ref{sec:cloud_stars}); and discuss one star identified as likely belonging to the Magellanic Bridge (Section \ref{sec:magellanic_bridge}).

\section{Data reduction and abundance determination}\label{sec:data_reduction}

The analysis presented in this work makes use of spectra obtained between 2013 November and 2017 September with the 3.9-metre Anglo-Australian Telescope at Siding Spring Observatory with the multi-fibre-fed HERMES spectrograph \citep{Sheinis2015} and the Two-Degree Field (2dF) top-end \citep{Lewis2002}. 2dF allows for the acquisition of up to 360 science targets per exposure when using 25 sky fibres. HERMES simultaneously acquires spectra using four independent cameras with non-contiguous wavelength coverage totalling $\sim1000$~\AA\ at a spectral resolution of $R\approx28,000$. Its fixed wavelength bands are 4715--4900~\AA, 5649--5873~\AA, 6478--6737~\AA, and 7585--7887~\AA, which were selected to cover absorption features from at least 29 chemical elements in giant and dwarf stars, sampling the major element groups and nucleosynthetic processes. The spectra were reduced using an \textsc{iraf}-based pipeline \citep{Tody1986,Tody1993}, which is described in detail in \citet{Kos2017}. Briefly, it performs initial quality checks, optimal extraction, reduction, and basic analysis of spectra, including determining the radial velocity.

The observations that make up the dataset described in this work come from three separate, but related, surveys: the GALAH pilot survey \citep{Duong2017}, the main GALAH survey \citep{DeSilva2015a}, and the \textit{TESS}-HERMES survey \citep{Sharma2017a}. These three programmes share infrastructure in terms of observing, data reduction, and abundance analysis, but have different aims and therefore different selection functions. In this work we concentrate on the 69,095 stars observed by any of these surveys within 15~deg of either MC, for which there was a match in UCAC5 \citep{Zacharias2017} with a proper motion error less than 2.5~mas\,yr$^{-1}$. This cutoff was selected because it included 99 per cent of the stars.

For both the main GALAH survey and the \textit{TESS}-HERMES survey, the selection functions were simple magnitude cuts. The input catalogue used is the union of 2MASS \citep{Skrutskie2006}, APASS \citep{Henden2016} and UCAC4 \citep{Zacharias2013} catalogues, with selections for photometric quality and crowding. Because of a lack of complete APASS photometry at the beginning of the GALAH survey, a $V$ magnitude for each star was estimated from the 2MASS $J,K_S$ magnitudes (Equation \ref{eq:mag}).

For the GALAH main survey, most fields have the magnitude range $12\leq V_{JK}\leq14$, with a smaller number of stars from fields with a magnitude range of $9\leq V_{JK}\leq12$. For the \textit{TESS}-HERMES survey the stars have $10\leq V_{JK}\leq13$. For the GALAH pilot survey, the selection function is more complicated, as these stars were chosen to investigate clusters, or the thin and thick disk of the Galaxy \citep[see][]{Duong2017}. For the 69,095 stars considered in this work, 85 per cent have $12\leq V_{JK}\leq14$. A total of 19,730 of the stars come from \textit{TESS}-HERMES, and 49,365 are from either the pilot or main GALAH surveys (for simplicity, we will group these two GALAH surveys together in the subsequent discussion).

We make use of labels that were determined using the methods described in \citet{Buder2018}. We note that we include some stars that are not found in either \textit{TESS}-HERMES DR1 \citep{Sharma2017a} or GALAH DR2 \citep{Buder2018}. One of the criteria used for those data releases was for the radial velocities measured for the blue, green, and red arms of HERMES be consistent. This was found to reject many stars located at the position of the MC. Inspection of their spectra identified that these were very hot stars, and the combination of the H-$\alpha$ emission and the strong He~I line at 6678~\AA, was leading to spurious radial velocities. As such, in this work we make use solely of the radial velocity determined from the blue camera spectra. This has a negligible offset of $0.06\pm0.90$~km\,s$^{-1}$ from the combined three-arm radial velocity normally used by GALAH or \textit{TESS}-HERMES.

\section{Photometric selection of likely Magellanic Cloud stars}\label{sec:initial_select}

In this section we develop criteria for identifying likely Magellanic Cloud (MC) supergiant stars; first, from SkyMapper+2MASS photometry (Section \ref{sec:skymapper_select}), and then using these to inform a label-based selection (Section \ref{sec:final_criteria}).

\subsection{SkyMapper colour selection of Magellanic Cloud stars}\label{sec:skymapper_select}
Before turning to the GALAH parameters, first we explore the photometry of the MC. In particular, we use the gravity-sensitive photometry of the SkyMapper Southern Sky survey \citep{Keller2007,Wolf2017} and 2MASS to identify likely MC stars, which will help inform the search of \textit{The Cannon} label space for these stars. SkyMapper is a 1.3-m telescope at Siding Spring Observatory that is undertaking a multi-epoch photometric survey of the whole southern sky in six photometric bands: $uvgriz$ \citep{Bessell2011}. The filter set of SkyMapper is superficially similar to that of Sloan Digital Sky survey \citep{Gunn1998} and Pan-STARRS \citep{Tonry2012}, but with some key differences. Relevant to this work is the addition of a narrow $v$ filter centred at 384~nm that is similar to the DDO 38 band. The $u$ and $v$ filters straddle the Balmer jump and so $u-v$ at a fixed colour (e.g., $g-i$) is gravity sensitive\footnote{The same regions of the spectrum also provide metallicity sensitivity for giants, see e.g., \citet{Keller2014,Jacobson2015,Howes2016,Simpson2018}.}. We can use this to distinguish blue AF-type main-sequence, horizontal-branch, and luminous blue supergiants stars of the same $g-i$. The unevolved red stars can be distinguished from the evolved stars of the same $g-i$ using a combination of SkyMapper and 2MASS photometry \citep[see figure 16 of][]{Wolf2017}.

We downloaded the SkyMapper DR1.1 \citep{Wolf2017} photometry for all stars within 15~deg of either MC that met the following global quality criteria:
$\mathrm{\texttt{nch\_max}} =1$ (source never resolved into multiple components) and $\mathrm{\texttt{class\_star > 0.95}}$ (highly likely to be a stellar target); and the per filter criteria of $\mathrm{\texttt{x\_good}} > 0$ (source detected in at least one image in all filters),
$\mathrm{\texttt{x\_nimaflags}} = 0$ (isophotal aperture clean of bad pixels, saturation, cross-talk, cosmic rays), and
$\mathrm{\texttt{x\_flags}} = 0$ (no Source Extractor warnings about saturation, close neighbours, edge-of-CCD effects, etc) where $\mathrm{\texttt{x}}=[\mathrm{\texttt{u}},\mathrm{\texttt{v}},\mathrm{\texttt{g}},\mathrm{\texttt{i}}]$. These stars were positionally cross-matched to UCAC5 stars with proper motions errors $<2.5$~mas\,yr$^{-1}$, and to 2MASS stars with a photometric quality of "A" for their $K_S$ magnitude. This gave a final catalogue of 724,940 stars, of which 418,952 had UCAC5 proper motions, and 622,933 have a $K_S$ magnitude. The proper motions were converted to the Galactic frame of reference using \textsc{astropy} \citep{TheAstropyCollaboration2018}.

We define an SMC region of the sky as those stars within 3.0~deg of $(\mathrm{RA},\mathrm{Dec})=(15.1, -73.0)$ and/or 2.5~deg of $(\mathrm{RA},\mathrm{Dec})=(28.6, -73.5)$; and an LMC region as those stars within 6.5~deg of $(\mathrm{RA},\mathrm{Dec})=(81.5, -68.5)$. Stars outside of these regions we will refer to as ``field'' stars, while stars within these regions are referred to as ``Magellanic Cloud'' stars, though obviously there will be contamination of Galactic stars into the MC regions, and vice versa. These regions are somewhat conservative: the LMC disc extends $\sim7$~deg from the centre \citep{Mackey2016}, with further evidence for kinematically related stars out to $\sim20$~deg \citep{Majewski1999,Majewski2008} and a stellar halo that extends to at least $\sim30$~deg \citep{Belokurov2016}. However, as will be shown in this work, the bulk of the stellar component of the MC that we observed is found within these bounds.

\begin{figure*}
    \includegraphics[width=\textwidth]{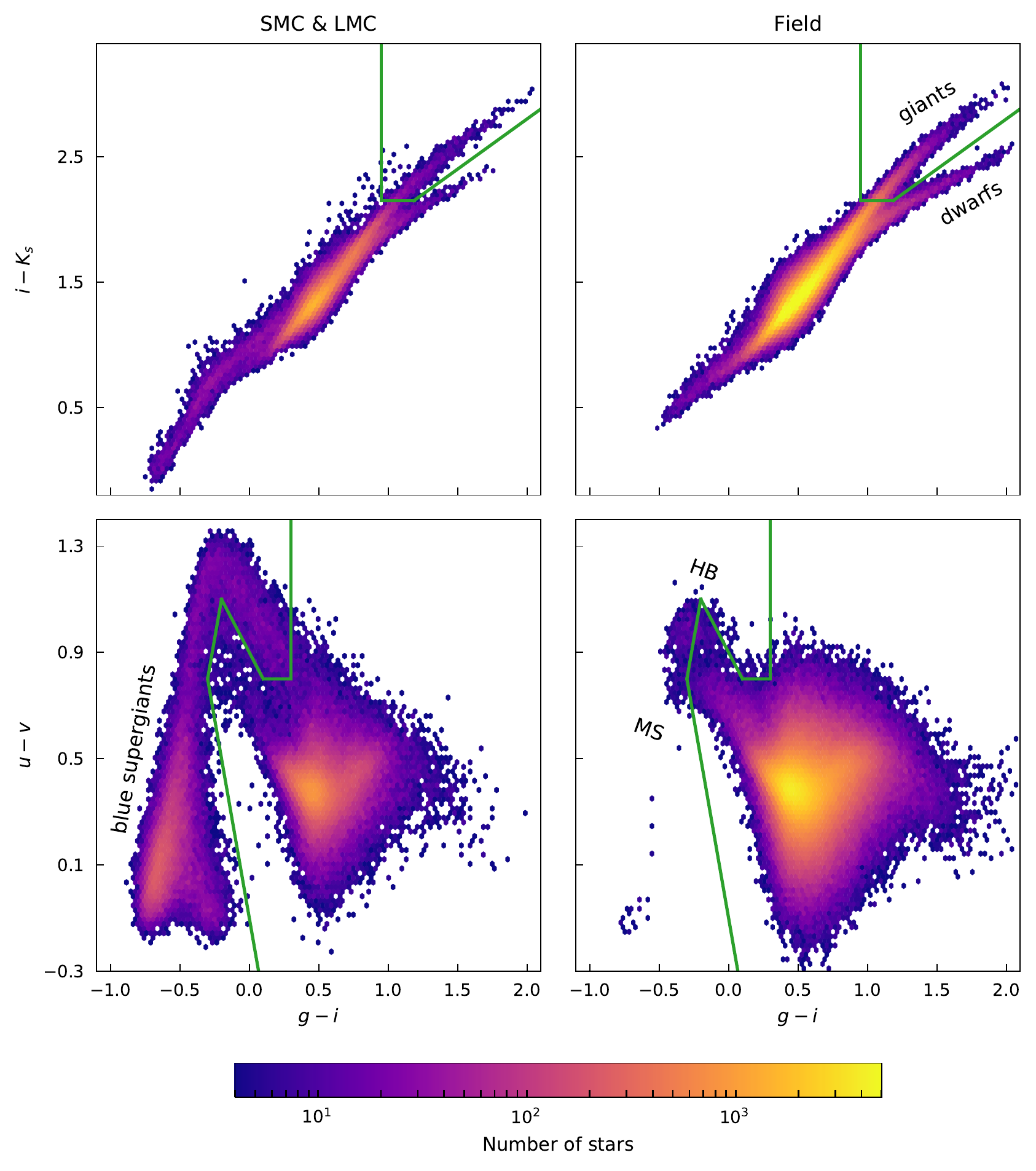}
    \caption{The distribution in colour-colour space of stars from the LMC and SMC regions (left column) compared to those from the field (right column). In the top row, the $i-K_S$ colour provides good separation of red giants and dwarfs, with the MC region have a much less populated lower sequence of red stars. In the bottom row, the $u-v$ colour shows a sequence of blue stars in the MC that is not very well populated in the Milky Way; these are the blue supergiants in the MC. The green lines show the cuts used to select the blue and red stars: in the top panels, stars above the green line were selected as possible red supergiants; and in the bottom panels, stars to the left of the green lines were selected as possible blue supergiants.}
    \label{fig:gz_smc_lmc}
\end{figure*}

\begin{figure}
    \includegraphics[width=\columnwidth]{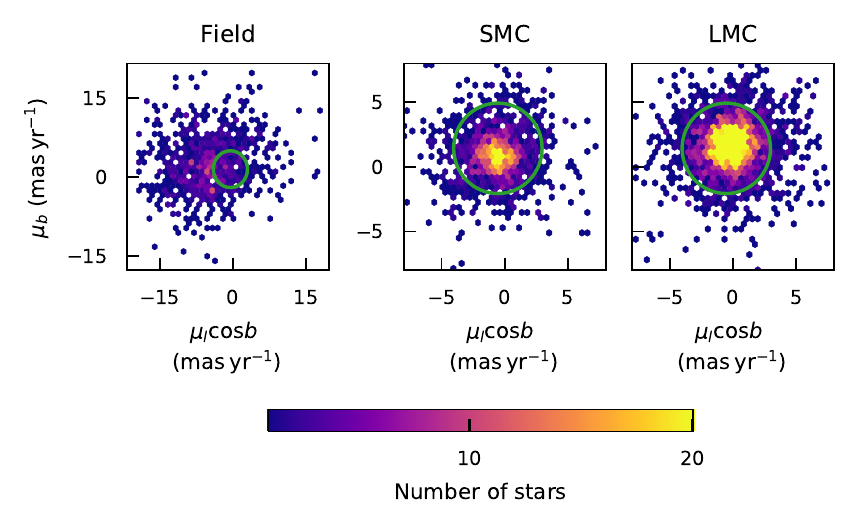}
    \caption{The proper motions distribution of the stars that met the blue supergiant colour criteria from the field, SMC, and LMC regions. The green circle on all the panels defines the proper motion selection for MC stars. For the MC regions, the stars are highly concentrated in proper motion, which would be expected for stars that are truly from these satellite galaxies, i.e., their proper motion is dominated by the motion of their host galaxy. There is a slight offset between the two MC, but a single proper motion selection is defined with the green circle that encompasses 86 per cent of the blue supergiants from the MC regions.}
    \label{fig:all_sm_pm}
\end{figure}

Figure \ref{fig:gz_smc_lmc} compares the colour-colour distributions of the MC (left column) and field regions (right column). In all the plots, $g-i$ is shown on the x-axis, and it is combined with two different colours: $i-K_S$ (top row) and $u-v$ (bottom row). The former is gravity sensitive for red stars, while the latter is gravity sensitive for blue stars. On the top row, the red stars bifurcate at $g-i>1.3$, with a population of red dwarf stars with lower $i-K_S$ that is found in the field, but is not nearly as populated in the MC regions. The  sequence with larger $i-K_S$ are the evolved red stars \citep[see also figure 16 of][]{Wolf2017}. Stars with $g-i > 0.95$, $i-K_S>2.3$ and $i-K_S > 0.8(g-i)+1.2$ were selected as likely evolved stars, but there will be some contamination of Galactic halo giants into this selection for the MC regions. This contamination is explored later in detail in Section \ref{sec:final_criteria} (Figure \ref{fig:red_stars}).

For the blue stars, the SkyMapper photometry is effective at distinguishing various stages of stellar evolution and mass. For stars with $g-i\sim-0.4$ and $u-v>0.7$, there are three populations of stars: two are primarily present in the field population (bottom-right panel), while a third can be found in the MC regions (bottom-left panel). This is distinguishing between evolved and unevolved stars: the population at $u-v\approx0.8$ are blue main sequence stars; $u-v\approx1.0$ are the horizontal branch stars; and the $u-v\approx1.2$ stars are a population of blue supergiant stars in the MC. For the MC region, these blue supergiants form a long sequence of stars that runs from $-0.2<u-v<1.4$, and then has a turn-over towards the main bulk of the stars. This population is present in the field regions but is nowhere near as populated. We define a selection of blue supergiants shown with the green lines on the bottom panels of Figure \ref{fig:gz_smc_lmc}.

Any star that belongs to either of MC should have very similar, but small, proper motions. Taking just the stars selected as likely blue supergiants (because they have the least contamination from Galactic stars) we show the proper motion distribution in Figure \ref{fig:all_sm_pm}. The MC blue stars are found within a small region of proper motion space compared to the field regions. Our SMC sample had an average proper motion of $(\mu_\mathrm{RA},\mu_\mathrm{Dec})=(0.9,-1.0)\pm(3.2,2.6)$~mas\,yr$^{-1}$, and our LMC sample had $(\mu_\mathrm{RA},\mu_\mathrm{Dec})= (1.7,-0.0)\pm(2.7,3.6)$~mas\,yr$^{-1}$ in UCAC5. It should be noted that these are a non-uniform spatial averages across a large field of view. The best determinations of the proper motion of the centre of mass of the SMC and LMC is that based on HST data by \citet{Kallivayalil2013}. They have done the necessary full treatment required, allowing for projection effects and rotation of the MC; something that is beyond the scope of this work (and beyond the uncertainties of the UCAC5 proper motions). Allowing for sign convention changes, our averaging process has produced mean values in excellent agreement with the values from \citet{Kallivayalil2013} given the uncertainties and simple approach.
 
Although the SMC and LMC do have slightly different proper motions, we proceed with a combined proper motion selection, because one of the aims of this work is to identify stars that belong to the Magellanic Bridge, not just the MC themselves. We consider likely MC stars to be found at a Cartesian distance of less than 3.5~mas\,yr$^{-1}$ from $(\mu_\mathrm{l}\cos b,\mu_\mathrm{b})=(-0.54, 1.44)$~mas\,yr$^{-1}$. This region is indicated with the green circles on Figure \ref{fig:all_sm_pm} and includes 86 per cent of the blue stars from the MC regions.

\begin{figure}
    \includegraphics[width=\columnwidth]{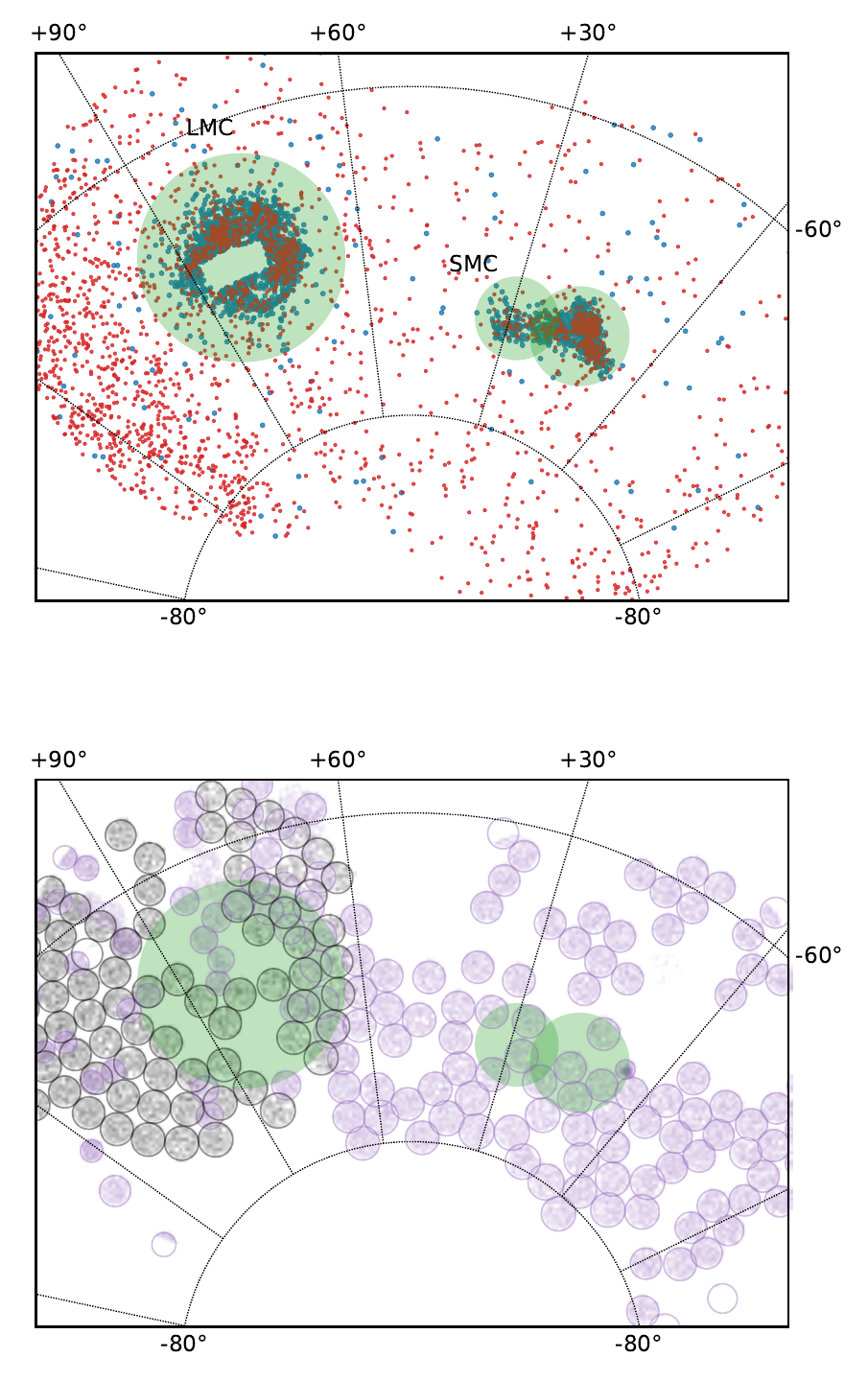}
    \caption{Top: Sky location of the stars observed by SkyMapper that meet the UCAC5 proper motion criteria and either of the blue (blue dots; using the $g-i$ and $u-v$ plane) and red SkyMapper+2MASS colour criteria (red dots; using $g-i$ and $i-K_s$). The green shaded regions indicate the SMC and LMC regions defined in this work. These criteria have identified stars that very much tend to be found at the locations of the MC. Stars outside of the MC regions, in particular the red stars, tend to be at higher RA, which is towards lower Galactic latitude. The missing rectangle in the middle of the LMC is the result of SkyMapper lacking reliable $u$ and $v$ measurements in those regions. Bottom: Location of the observed stars; fields with black edges are from the \textit{TESS}-HERMES survey, purple is for regular GALAH survey fields. There are no fields observed with declinations $<-80$~deg. The over-density of GALAH observations at the right of the SMC is the globular cluster 47 Tuc.}
    \label{fig:all_stars_map_galah}
\end{figure}

The overall sky distribution of all stars within 15~deg of either MC that met the proper motion criteria and either of the blue or red colour criteria is shown in the top panel of Figure \ref{fig:all_stars_map_galah}. These stars are strongly concentrated at the position of the MC on the sky, and the red stars outside of the MC regions tend to be found to the left of the top panel of Figure \ref{fig:all_stars_map_galah}, which is the direction of lower Galactic latitude. In the bottom panel we show the distribution of stars observed by GALAH and \textit{TESS}-HERMES. The observing coverage over both MC and the surrounding regions is uneven, especially to the north of the SMC, but each MC has several fields covering it. Of the 69,095 stars observed by either GALAH or \textit{TESS}-HERMES, 9720 are in the LMC region and 3132 are in the SMC region.

In the following section we use these colour-based criteria to investigate the location in \textit{The Cannon} label space that these MC supergiants are found.

\subsection{Cannon labels of Magellanic Cloud stars}\label{sec:final_criteria}
\begin{figure}
    \includegraphics[width=\columnwidth]{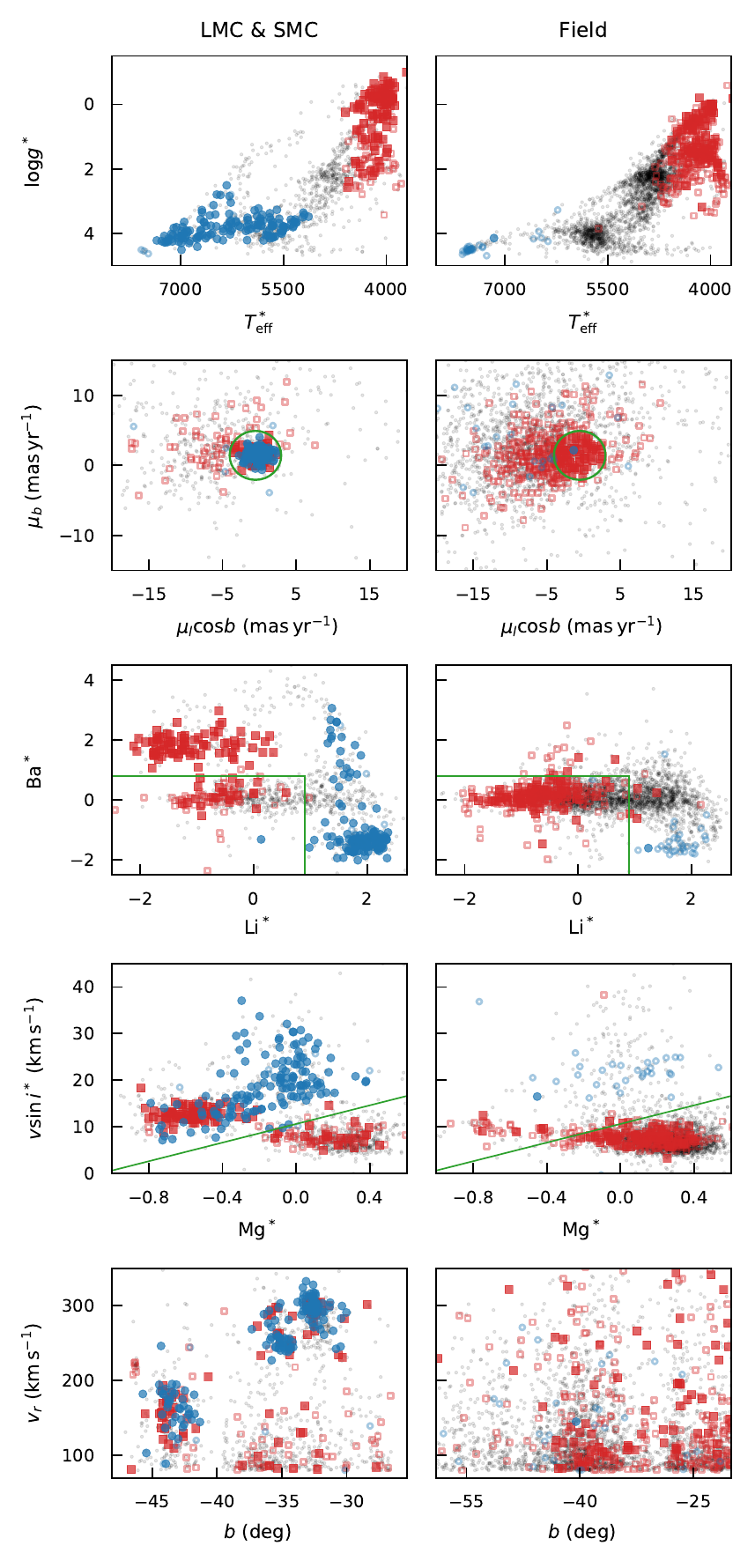}
    \caption{Stars observed by GALAH or \textit{TESS}-HERMES with radial velocities $v_r>80$~km\,s$^{-1}$ that met the SkyMapper blue (circles) and red (squares) colour criteria. The green lines indicate the criteria used in Section \ref{sec:final_criteria} to identify MC supergiants. Filled symbols are used for stars inside the proper motion circle on the panels on the second row, and open symbols for those outside of it. The left column shows stars from the LMC \& SMC regions, and the right column are those for the `field' region. Inspection of \textit{The Cannon} label space identified \life, \bafe, \mgfe, and \vsinidag\ as being the most helpful at distinguishing these blue and red MC stars from the rest of the stars. The MC red stars have larger \bafe\ than the field stars, and the blue stars tend to have high \life.}
    \label{fig:mg_ba_vsini_colour}
\end{figure}

In theory it should be possible to simply use the stellar labels (e.g., \teff, \logg, \feh\ etc.) to identify the MC supergiants. OB supergiants have $10000<\teff<50000$~K and $\logg\sim3$ \citep[e.g.,][]{Massey2005,Bouret2012}, while the cool red supergiants will have very low temperatures and surface gravities, like Betelgeuse \citep[$\teff=3500$~K and $\logg=-0.5$;][]{Lobel2000}, or Antares \citep[$\teff=3660$~K and $\logg=-0.2$;][]{Ohnaka2013}. The metallicities of the MC are well-known \citep[e.g.,][]{Russell1992,Rolleston2002,Dobbie2014}. But the training set used by the GALAH and \textit{TESS}-HERMES implementation of \textit{The Cannon} limits the outputs to a range of $3700\mathrm{~K}<\teff<7800\mathrm{~K}$ and $-0.6<\logg<5.0$, which is where the stars of interest to the main aims of the surveys are found. It is not obvious that the hottest stars will `pile up' at the edges of the label space. \citet{Casey2016a} found that when applying \textit{The Cannon} to RAVE spectra the hot stars outside of the training set bounds project into a single clump in the label space near the turn-off.

Before continuing, it is important to note that \textit{for these particular luminous MC stars} we are not discussing the labels with a belief that they are necessarily accurate. As will be shown, labels like \teff, \logg, \feh, \lireal, \bareal\ are erroneous for these stars because their parameters are well outside the bounds of the training set, and do not agree with the literature. However, because of the data-driven nature of \textit{The Cannon}, there is clearly coherence in their positions in the label space, and it is this coherence that we exploit in this work. We are ignoring all the flags in the dataset designed to remove these unreliable values, so, to avoid confusion, we will refer to all labels with a superscript $^\star$, e.g., \teffdag, \life, instead of \teff, \lireal, to denote that their absolute values are not to be considered accurate.  

In Figure \ref{fig:mg_ba_vsini_colour} we use the colour selections from Section \ref{sec:skymapper_select} to show where the blue and red supergiant stars with radial velocities larger than $80$~km\,s$^{-1}$ are found in the label space. The requirement of $v_r>80$~km\,s$^{-1}$ is designed to just capture the lowest $v_r$ stars of the SMC. In the top row is the \teffdag-\loggdag\ distribution where we find that, like \citet{Casey2016a}, the blue supergiants from the MC regions are incorrectly assigned temperatures in the range $5000\mathrm{~K}<\teffdag<8000$~K and surface gravities $\loggdag\sim4$.

The \teffdag\ returned by \textit{The Cannon} for the blue stars are actually anti-correlated with their expected temperatures: for these blue stars, the coolest stars as found by \textit{The Cannon} are actually the hottest O stars. We are hitting the hot \teff\ edge of \textit{The Cannon}'s training set. Note that the \loggdag\ is forced higher because this is the only region \textit{The Cannon} has training set with high \teff. But due the grid effect, now the \teffdag\ is grossly underestimated, and hence the \bafe\ is underestimated as a result. The grid effect on \teffdag\ also forces \textit{The Cannon} to choose a higher \vsinidag\ to further mimic the ``featurelessness'' that is still not accounted by these \teffdag.

The red supergiants have ended up in their mostly expected place in the \teffdag-\loggdag\ space, with cool temperatures and low surface gravities. These cool supergiants are slightly cooler than \textit{The Cannon} coolest training sample, so \textit{The Cannon} can still find reasonable \teffdag-\loggdag\ for them. However, due to extrapolation, the \teffdag\ estimated is still warmer than the \teff\ truth. And because the \teffdag\ is too warm compared to the truth, this pushes up the \bafe\ (and other abundances).

The distributions of the blue and red stars in the field regions are different from those in the MC regions: in particular, the red stars are found at higher \loggdag. As would be expected, based upon the results from the full SkyMapper dataset, the MC supergiants are mostly concentrated into a small region of proper motion space. Those stars that are outside the proper motion region but meet the colour criteria are indicated with open symbols.

The aim of this work is to develop a \textit{Cannon} label-based method (supplemented with radial velocities and proper motions) for identifying MC stars, without knowledge of the photometry of the star. Not every star observed by GALAH or \textit{TESS}-HERMES has reliable SkyMapper photometry, especially in the gravity sensitive $u$ and $v$ filters. And solely using our colour selection does not provide a clean sample. In the middle row of Figure \ref{fig:mg_ba_vsini_colour} the \bafe\ label for the MC regions shows a bimodal distribution for the red stars; the bulk have $\bafe\sim2$, while there are some with $\bafe\sim0$. In the field regions, almost all of the red stars are found to have $\bafe\sim0$.

\begin{figure}
    \includegraphics[width=\columnwidth]{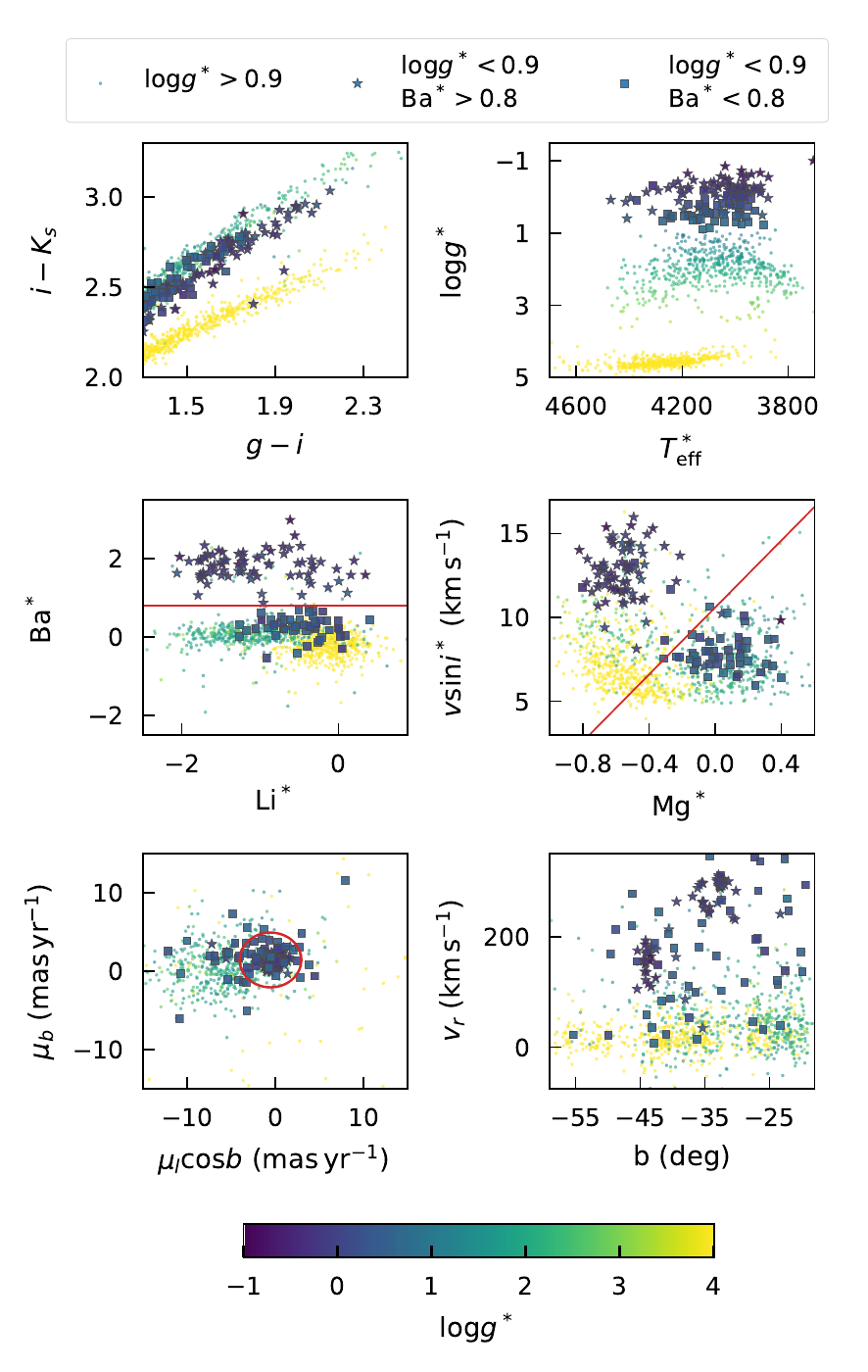}
    \caption{The reddest stars ($g-i>1.3$) observed by GALAH or \textit{TESS}-HERMES are shown in various parameter spaces. In all panels, the stars are coloured by their \loggdag. Stars with $\loggdag>0.9$ (Galactic dwarfs and giants) are shown with small symbols. The low gravity ($\loggdag>0.9$) targets are split with their \bafe: star symbols for $\bafe>0.8$ (likely MC stars) and squares for $\bafe<0.8$ (likely Galactic stars). The horizontal line in the middle-left panel indicates this \bafe\ split. The circle in the bottom-left panel is the proper motion selection defined in Section \ref{sec:initial_select} (Figure \ref{fig:all_sm_pm}). As observed in Figure \ref{fig:mg_ba_vsini_colour}, these \bafe-high stars are the likely MC red supergiants and this is confirmed by the concentration of these stars at the radial velocity and Galactic latitude of the MC.}
    \label{fig:red_stars}
\end{figure}

\begin{figure}
    \includegraphics[width=\columnwidth]{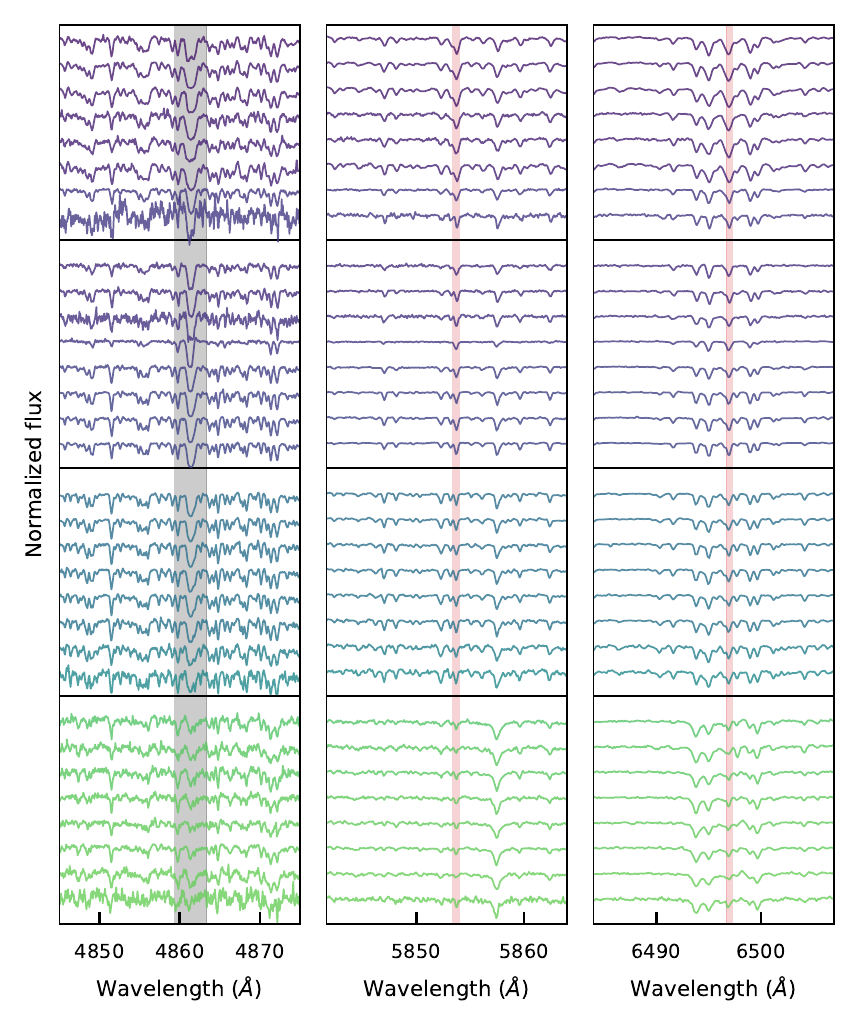}
    \caption{Portions of the normalized spectra of stars with $1.3<g-i<1.9$ from the blue (left), green (middle) and red (right) cameras of HERMES. Highlighted with the red-shaded regions are the barium lines at 5854~\AA\ and 6497~\AA\ which are used by \textit{The Cannon} to determine \bafe. The left panel shows the H-$\beta$ line confirming that these stars do have different gravities. The top section of spectra in each panel are stars with $\loggdag<0.9$ and $\bafe>0.9$ (star symbols on Figure \ref{fig:red_stars}). The next section spectra are examples stars with $\logg<0.9$ and $\bafe<0.9$ (circles on Figure \ref{fig:red_stars}). The next section has $0.9\geq\loggdag<3.7$ (giants), and the bottom section has stars with $\loggdag\geq3.7$.}
    \label{fig:spec_stacked_giants}
\end{figure}

In Figure \ref{fig:red_stars} we explore the label distribution of the red stars further. Plotted are all stars (not just those with $v_r>80$~km\,s$^{-1}$) observed by GALAH or \textit{TESS}-HERMES within 15~deg of either MC with $g-i>1.3$, the colour at which the colour-colour diagram bifurcates between evolved and unevolved stars (see also Figure \ref{fig:gz_smc_lmc}). \textit{The Cannon} is very successful at distinguishing these two categories of stars, but as the \teffdag-\loggdag\ diagram shows, there is a large range of surface gravities of the evolved stars ($-1<\loggdag<3$). The lowest gravity stars have the bimodal distribution of \bafe\ also found in Figure \ref{fig:mg_ba_vsini_colour}, and inspection of their spectra (Figure \ref{fig:spec_stacked_giants}), confirms that there is a difference in the strength of their barium lines. The \bafe-high stars are found to concentrate at the position, proper motions, and radial velocities of the MC, while the barium-normal population are found at a range of radial velocities and are scattered across the sky. They also show a difference in their \mgfe\ and \vsinidag\ labels: the \bafe-high stars have low \mgfe\ but large \vsinidag, and form a distinct group from the other stars with $\loggdag<0.9$.

Returning to Figure \ref{fig:mg_ba_vsini_colour}, since the aim is to identify all supergiants in the MC and their surroundings, it is not optimal to simply use \bafe, because this would exclude the blue stars, which are found at a large range of \bafe. Exploration of the label space identified a combination of \life, \bafe, \mgfe, and \vsinidag\ as being very good at differentiating MC supergiants from the rest of the sample. As already shown, the red stars have high \bafe\ values. We also find that the blue stars have high \life, but there is some contamination of Galactic stars in the regime of $\life\sim1.8,\bafe\sim0$. This is removed using the \mgfe\ and \vsinidag\ values: the MC stars tend to have large \vsinidag\ for their \mgfe.

We reiterate that these labels are unreliable in terms of their use for abundance-based studies. For instance, the red supergiants in the MC have a barium abundance from \textit{The Cannon} of $\bafe\sim2$~dex, but such values have never been observed in the MC. \citet{VanderSwaelmen2013} observed a large sample of LMC disk and bar stars ($N=164$), finding a typical range of $0.0<\bareal<0.9$. Several clusters in the LMC were observed by \citet{Colucci2012} who found $\bareal\sim+0.9$ for individual stars in young clusters like NGC~1978 and NGC~1866. For the blue supergiants observed by GALAH, the \teffdag\ and \loggdag\ are obviously erroneous, so we can also assume their abundances would be too. Inspection of the spectra show there is no lithium line present in these stars, and \textit{The Cannon} has correctly flagged these as having no significant detection of the line.

Combining \textit{The Cannon} labels, radial velocity, and proper motion information for the stars, we create the following criteria for identifying MC supergiant stars observed by GALAH or \textit{TESS}-HERMES:
\begin{itemize}
	\item Radial velocity: $v_r>80$~km\,s$^{-1}$
	\item Proper motion: within $3.5$~mas\,yr$^{-1}$ of $(\mu_\mathrm{{l}\cos b},\mu_\mathrm{{b}})=(-0.54,1.44)$
	\item \textit{Cannon} labels: $10\times\mgfe+10.6 < \vsinidag$ and either $\bafe > 0.8$ or $\life > 0.9$.
\end{itemize}

In the following section we apply these criteria to all of the stars observed by GALAH within 15~deg of either MC and find them to be extremely successful, with a low rate of false positives or negatives, at identifying supergiant stars in and around the MC.

\section{Magellanic Cloud stars selected from \textit{Cannon} labels}\label{sec:cloud_stars}

\begin{figure*}
    \includegraphics[width=\textwidth]{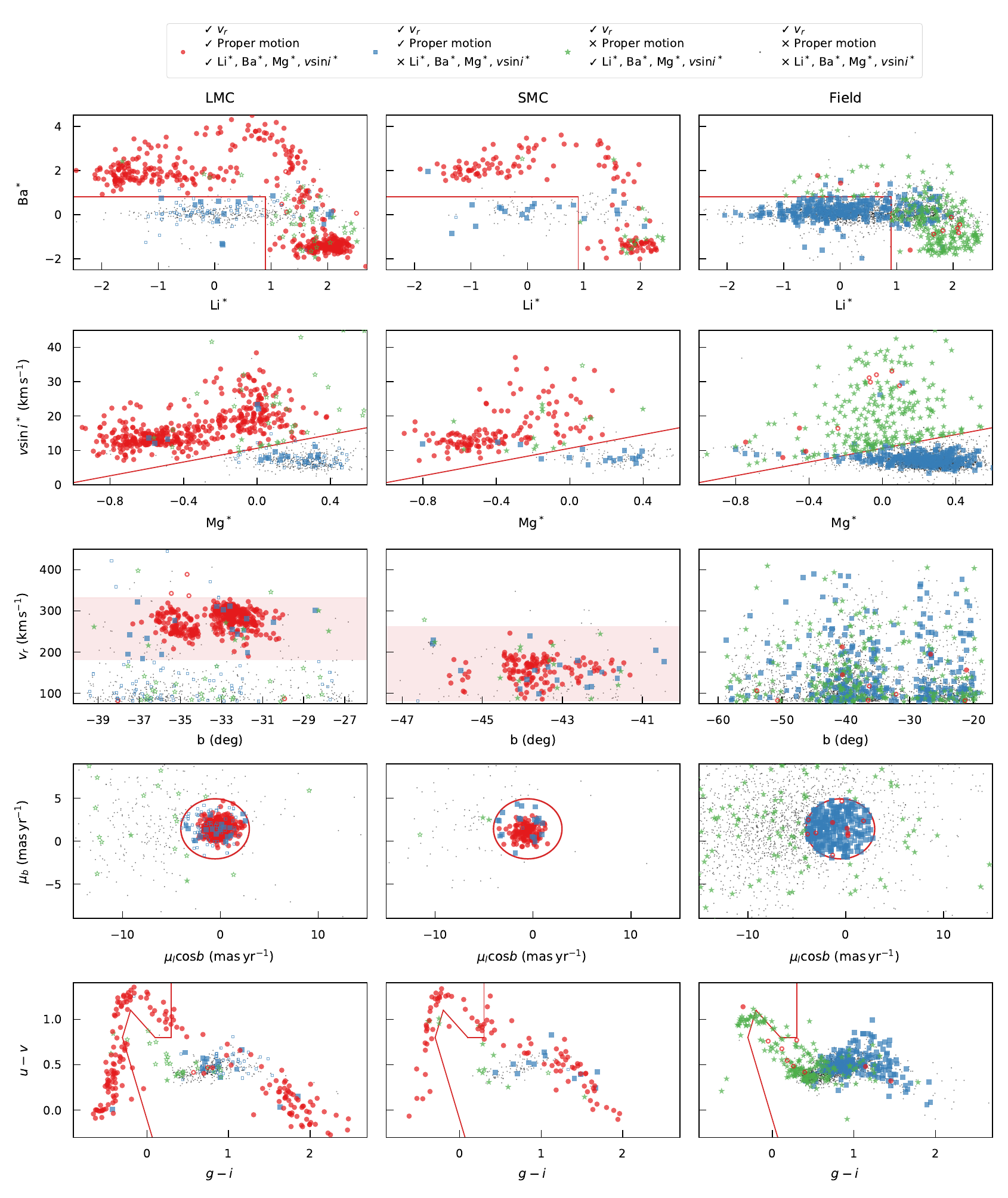}
    \caption{The supergiant stars of the LMC, SMC and surrounding field regions were identified as those with high enough radial velocity, within the proper motion bounds of the MC, and within certain bounds for their label values. The stars have been placed into four categories: stars that meet all the three criteria (red dots); stars that fail the label criteria (blue squares); stars that only fail the proper motion criteria (green stars); stars that fail both the label and proper motion criteria (small black dots). For the LMC and SMC regions, filled symbols are used for stars within the expected radial velocity range. For the field region, open symbols are used for false positives which were rejected due to their photometry.}
    \label{fig:mg_ba_vsini}
\end{figure*}

\begin{landscape}
\begin{table}
\caption{IDs, positions, and \textit{Cannon} labels for the 581 stars identified as MC supergiants. The full version is available from the journal.}\label{table:stars}
\begin{tabular}{rlrlrrrrrrrrrrrrlr}

\hline
     sobject\_id &              smss\_j &           gaia\_source\_id &       GCVS &       RA ($\degr$) &       Dec ($\degr$) & $v_r$ (km\,s$^{-1}$) & 
     \teffdag\ (K) & \loggdag & \fehdag & \vsinidag & \life & \mgfe & \bafe & VarType & Period (days) \\ 
\hline
150107002201317 &  053954.29-672319.6 & 4659471996367683584 &            & 84.97625 & -67.38881 &       291.8 &        7277 &        4.24 &      -0.46 &         14.4 &            1.52 &           -0.35 &           -1.39 &         &        \\
140807005601298 &  010502.33-725532.6 & 4687406669826115584 &  SMC V1912 & 16.25962 & -72.92569 &       193.3 &        4158 &       -0.47 &      -0.42 &         12.4 &           -1.28 &           -0.60 &            2.12 &     LC: &        \\
140806004701098 &  005447.81-712452.9 & 4689300200660147200 &            & 13.69929 & -71.41469 &       182.1 &        6990 &        4.30 &      -0.63 &         27.9 &            1.77 &           -0.27 &           -1.32 &         &        \\
161217003601102 &                     &                     &            & 75.22979 & -66.63731 &       319.4 &        5246 &        0.85 &      -0.01 &         16.1 &           -0.15 &           -0.46 &            3.93 &         &        \\
161107003901091 &  053040.67-672308.0 & 4660172969384724480 &            & 82.66925 & -67.38556 &       304.8 &        4047 &       -0.37 &      -0.14 &         13.0 &           -1.70 &           -0.73 &            2.02 &         &        \\
161217003601069 &                     &                     &            & 75.84304 & -66.36983 &       294.9 &        5996 &        1.86 &       0.16 &         12.7 &            0.88 &           -0.62 &            3.37 &         &        \\
161116002201011 &  004348.44-733648.7 & 4685823716674504704 &  SMC V0284 & 10.95183 & -73.61356 &        86.1 &        6082 &        2.19 &      -0.04 &         15.8 &            1.25 &           -0.84 &            3.16 &    DCEP &  31.93 \\
170115002201207 &                     &                     &            & 73.81692 & -69.32006 &       245.2 &        3997 &        0.46 &      -0.39 &         13.7 &           -2.08 &           -0.44 &            2.01 &         &        \\
170115002201358 &                     &                     &            & 77.28621 & -68.98539 &       263.8 &        6127 &        1.97 &       0.26 &         14.3 &            0.98 &           -0.86 &            3.46 &         &        \\
170108002201380 &                     &                     &  LMC V3090 & 82.76425 & -69.09444 &       273.1 &        4012 &       -0.27 &      -0.12 &         13.1 &           -1.14 &           -0.68 &            2.04 &     LC: &        \\
170115002201195 &  045541.84-692624.2 & 4655173833609707520 &  LMC V0292 & 73.92433 & -69.44006 &       252.4 &        4195 &        1.08 &      -0.07 &         14.0 &           -1.07 &           -0.55 &            2.08 &     SRC & 675.00 \\
150107002201252 &                     &                     &            & 83.40013 & -67.50000 &       297.5 &        6435 &        2.87 &       0.07 &          9.8 &            1.33 &           -0.67 &            2.36 &         &        \\
161107003901127 &                     &                     &            & 82.19579 & -67.15900 &       301.2 &        5655 &        3.77 &      -1.54 &         22.0 &            1.89 &            0.16 &           -1.52 &         &        \\
150107002201069 &                     &                     &            & 85.41704 & -68.60675 &       284.0 &        7171 &        4.11 &      -0.51 &         11.7 &            1.69 &           -0.42 &           -1.55 &         &        \\
170115002201170 &                     &                     &            & 75.32654 & -69.54733 &       261.8 &        4100 &       -0.86 &      -0.04 &         13.7 &           -1.76 &           -0.76 &            2.95 &         &        \\
140807005601359 &  011235.17-730935.6 & 4687177387283550208 &            & 18.14654 & -73.15986 &       172.1 &        4196 &       -0.08 &      -1.11 &         13.7 &           -0.34 &           -0.21 &            1.98 &         &        \\
170115002201198 &  045329.41-692432.7 & 4655363946042099712 &            & 73.37246 & -69.40914 &       280.5 &        5507 &        3.91 &      -1.71 &         24.7 &            0.98 &            0.01 &           -1.59 &         &        \\
161107003901123 &                     &                     &            & 82.09154 & -67.29669 &       295.8 &        6062 &        3.53 &      -1.25 &         16.7 &            2.12 &           -0.17 &           -1.79 &         &        \\
161217003601020 &  050714.17-660318.5 & 4662061311885050880 &            & 76.80908 & -66.05511 &       282.5 &        6283 &        2.16 &       0.17 &         22.3 &            1.26 &           -0.72 &            2.93 &         &        \\
161217003601054 &                     &                     &            & 75.45875 & -66.13011 &       295.3 &        6857 &        3.48 &      -0.11 &         12.4 &            1.66 &           -0.70 &            1.09 &         &        \\
140807005601222 &                     &                     &            & 13.40058 & -73.23897 &       120.0 &        6831 &        4.37 &      -0.75 &         33.3 &            1.92 &            0.13 &           -1.05 &         &        \\
161107003901397 &  053334.37-665109.9 & 4660260621076648960 &            & 83.39321 & -66.85278 &       314.0 &        5544 &        3.99 &      -1.64 &         26.1 &            1.90 &            0.01 &           -1.49 &         &        \\
170115002201276 &  045943.53-683122.8 & 4661327319154037760 &            & 74.93142 & -68.52300 &       276.1 &        6226 &        1.93 &       0.34 &         16.8 &            1.01 &           -0.85 &            3.61 &         &        \\
140807005601330 &                     &                     &            & 17.62279 & -72.82628 &       134.6 &        4475 &        0.06 &      -0.51 &         12.0 &            0.46 &           -0.47 &            2.20 &         &        \\
170108002201148 &                     &                     &            & 79.96583 & -69.88569 &       257.3 &        5279 &        3.49 &      -1.82 &         18.3 &            1.15 &           -0.28 &           -1.06 &         &        \\
170108002201166 &                     &                     &            & 80.41142 & -69.47181 &       270.4 &        5431 &        3.34 &      -1.69 &         15.2 &            1.91 &           -0.04 &           -1.35 &         &        \\
161115003201358 &  050228.82-704106.0 & 4655039349596015616 &            & 75.62013 & -70.68500 &       241.3 &        5502 &        3.77 &      -1.68 &         21.8 &            1.38 &            0.05 &           -1.22 &         &        \\
161107003901238 &                     &                     &            & 81.87917 & -66.58258 &       278.9 &        7109 &        4.34 &      -0.54 &         26.0 &            1.75 &            0.09 &           -1.46 &         &        \\
161219004101315 &  055600.34-674113.5 & 4659180213470700544 &            & 89.00137 & -67.68708 &       282.8 &        4111 &       -0.22 &      -0.27 &         14.4 &           -1.46 &           -0.62 &            1.96 &         &        \\
161107003901054 &                     &                     &            & 83.07508 & -67.05122 &       260.0 &        4016 &       -0.38 &      -0.21 &         11.5 &           -0.18 &           -0.72 &            2.07 &         &        \\
\hline
\end{tabular}
\end{table}
\end{landscape}

In this section we apply the proper motion, radial velocity, and \textit{Cannon} label selections to 69,095 stars observed by GALAH and \textit{TESS}-HERMES in and around the Magellanic Clouds (MC), with the aim of identifying members of MC themselves, as well as stars that are part of the surrounding structures of the Bridge and streams. We identify 581 possible MC supergiants. The full list of stars is in Table \ref{table:stars}. Figure \ref{fig:mg_ba_vsini} presents the parameters for three regions of the sky: the LMC (left column), SMC (middle column), and field regions (right column). The stars have been placed into four categories: stars that meet all the three criteria; stars that only fail the proper motion criteria; stars that fail the label criteria; stars that fail both the label and proper motion criteria.

In the field region there are only 10 stars that meet the criteria to be MC supergiants. There were 273 stars that failed the proper motion criteria, and 326 stars that failed the label criteria, and these two groups had different distributions in the parameter space. The former stars tend to be bluer, with a high lithium label. This is because of the relative lack of \bafe-rich stars in the field, so all of the stars that pass the label criteria are found to be \life-rich. This splitting of these two groups of stars manifests in the colour-colour spaces, where the lithium-rich stars have $g-i<0.8$. In the $u-v$ colour, there are two parallel sequences of stars: blue dwarfs and blue evolved stars. We will discuss false positives in Section \ref{sec:false_positives}, but we note here that six of the possible MC-like stars in the field are found on these Galactic star sequences. 

In the MC regions, we find a large fraction of stars identified as members of the MC. Of the 1013 stars observed in the LMC region with $v_r>80$~km\,s$^{-1}$, 434 stars were identified as members, 41 stars were rejected due to their proper motions, and 83 by their labels. For the SMC, there were 300 stars observed with $v_r>80$~km\,s$^{-1}$, with 137 identified as members, 16 rejected for their proper motions, and 21 for their labels. The rejected stars follow the same patterns observed in the field region, namely that the proper motion outliers are bluer, and the label outliers are redder. In Sections \ref{sec:false_positives} and \ref{sec:false_negatives} we discuss the false positives and negatives, in particular via the radial velocity and colour distributions. Here we highlight that in the MC regions almost every star identified as a supergiant falls within the literature radial velocity bounds of each MC (shaded red bands on the third row of Figure \ref{fig:mg_ba_vsini}).

By using a label selection rather than a colour selection, we have been able to identify MC stars in the colour range $0<g-i<1.5$, which were either too blue for the red colour selection, or vice versa. It is in this colour range that we find many of the stars that were rejected, in particular those stars that failed both the proper motion and label criteria, but still have $v_r>80$~km\,s$^{-1}$.

\subsection{General properties of Magellanic Cloud stars}

\begin{figure}
    \includegraphics[width=\columnwidth]{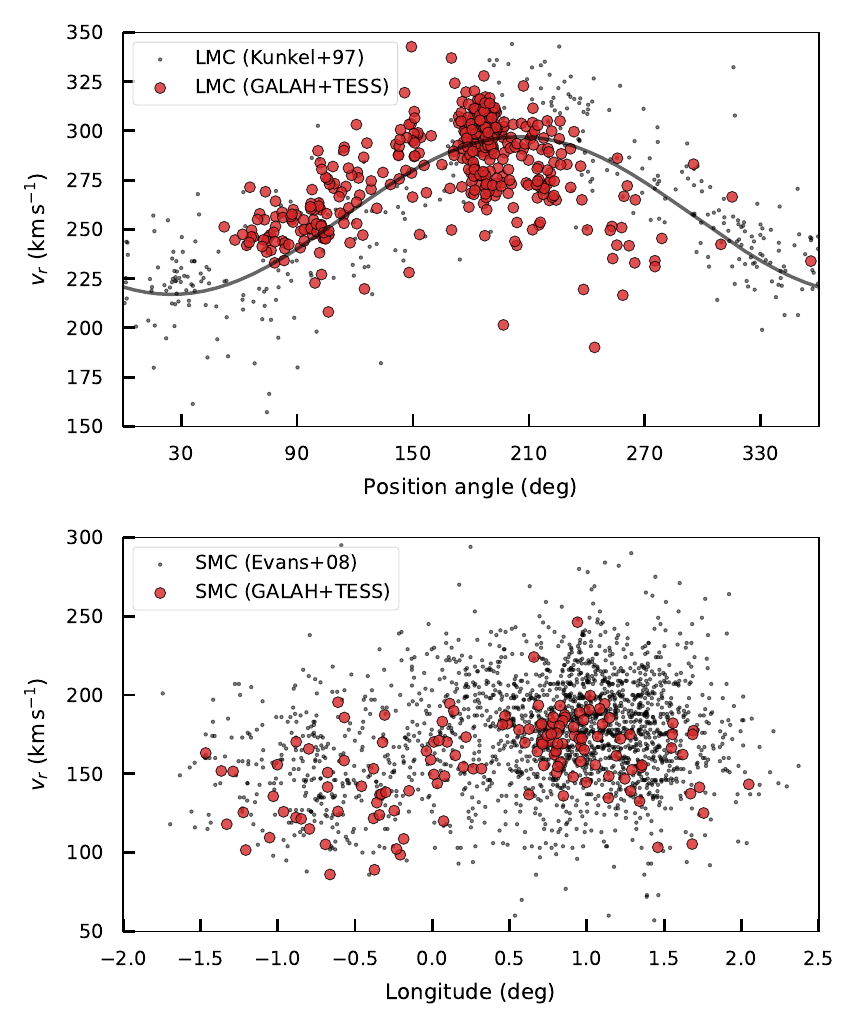}
    \caption{Radial velocity distribution of stars identified as MC members. Top: LMC stars observed by GALAH and \textit{TESS}-HERMES with respect to their position angle around the centre of mass as defined by \citet{VanderMarel2002}. The observed stars happened to sample the higher velocity stars so our average velocity is larger than the literature value of the LMC. The curve is representative and has been fitted by eye. Bottom: The SMC radial velocity with respect to their longitude along a line centered at $\mathrm{RA}=12^\circ, \mathrm{Dec}=-73^\circ$, with a position angle of $-50$~deg. It is compared to the distribution of radial velocities from \citet{Evans2008}.}
    \label{fig:rv_posangle}
\end{figure}

From the LMC region there are 434 members identified with an average radial velocity of $282\pm26$~km\,s$^{-1}$, and from the SMC there are 137 members identified with an average radial velocity of $158\pm27$~km\,s$^{-1}$. At first impressions, these values do differ from the literature systemic velocities of the MC, but this can be explained as being due to our uneven sampling of each MC and the complicated velocity distributions of the MC.

The plane of the LMC disk is tilted by about $35$~deg with respect to the plane of the sky \citep{VanderMarel2001,Olsen2002,Nikolaev2004} and so at different position angles the rotation velocity has a different component of the rotation velocity projected into the line of sight. As shown in the top panel of Figure \ref{fig:rv_posangle} the spectroscopically-observed stars are found at a small range of position angles and this resulted in a higher average velocity than the literature systemic velocity \citep[$\sim260$~km\,s$^{-1}$;][]{VanderMarel2002,Cole2004}.

For the SMC there are a range of reported values in the literature depending on the type of star and region observed: e.g., $146\pm28$~km\,s$^{-1}$ for red giant stars \citep{Harris2006}; $172\pm28$~km\,s$^{-1}$ for OBA type stars \citep{Evans2008}. The line-of-sight velocity distribution of the SMC is more complicated than the LMC and the uneven sampling of the SMC means we do not have a completely representative sample. Most of the stars observed on the SMC bar are from the southern end, and the rest of the stars come from the eastern wing. The bottom panel of Figure \ref{fig:rv_posangle} gives the radial velocity with respect to a longitude along the bar, ie., a line position centred at $\mathrm{RA}=12^\circ, \mathrm{Dec}=-73^\circ$, with a position angle of $-50$~deg. Due to our spatial sampling, we are missing the higher velocity stars that would increase our average to be more towards that of \citet{Evans2008}.

Although the metallicity of the stars has been flagged as unreliable by \textit{The Cannon}, we have still been able to recover that there are metallicity differences between the two MC. Considering only the red supergiants (defined as MC stars with $\life<0$), for which the \teffdag\ and \loggdag\ labels are more reasonable, we find the SMC has a mean metallicity of $\fehdag=-0.5\pm0.2$ and for the LMC $\fehdag=-0.2\pm0.2$. As with the radial velocities, these values are somewhat different from those found with dedicated studies: for the SMC values range from $[\mathrm{M}/\mathrm{H}]=-0.6$ for young stars \citep{Russell1992} to $\feh=-1$ for red giants \citep{Dobbie2014}; for the LMC, massive main sequence stars have been found to have $[\mathrm{Z}/\mathrm{H}]=-0.3$ \citep{Rolleston2002}. Due to the reasons mentioned before we will not be entering into discussions about the relative abundances of the two MC or comparing them to the Milky Way Galaxy. However as noted above, \textit{The Cannon} \feh\ label is not too far from the literature expectations. and moreover, the difference between the LMC and SMC \fehdag\ values is completely consistent with the observed literature \feh\ difference.

\subsection{False positives}\label{sec:false_positives}
In this and the following subsections, we consider two important questions: What is the rate of false positives (Section \ref{sec:false_positives}) for this method, and what is the rate of false negatives (Section \ref{sec:false_negatives})?

Here we explore two avenues for identifying false positives: the radial velocities of the stars selected as MC supergiants; and their colour-colour distributions. Both have their advantages and disadvantages. For the radial velocity, it is easy to reject a star that has a radial velocity that is much larger than the expected range for a given MC. But in the case of the SMC, the radial velocity is within the expected bounds of stars within the Milky Way (compare the radial velocity distribution of the SMC to the field regions in Figure \ref{fig:mg_ba_vsini}). Even the LMC has some overlap with the Milky Way velocity distribution. The weakness of the colour-colour distributions is that, although we know that the blue MC supergiants are found in a relatively unique part of the $u-v$/$g-i$ colour-colour space, the red stars have a large amount of confusion with evolved stars in the Milky Way (see e.g., Figure \ref{fig:red_stars}). And not every star observed by GALAH and \textit{TESS}-HERMES has reliable SkyMapper photometry in the useful filters.

Because of its relatively large radial velocity with respect to the Milky Way, the LMC provides a better testing ground for identifying false positives. There are only three stars with a radial velocity well outside the expected bounds: two with velocity $v_r<88$~km\,s$^{-1}$, and one with $v_r=389$~km\,s$^{-1}$ (likely a Galactic halo star; \textit{Gaia} DR1 4760338822280336256). Both of the low velocity stars are found in the $u-v$/$g-i$ colour-colour diagram at $g-i\approx0.7$ and $u-v\approx0.45$, a realm dominated by stars that are not MC stars. It is here that we find most of the stars that failed one of the criteria. There are another two stars identified as LMC supergiants that are found in this region of the colour-colour diagram, but these two stars have radial velocities much more like the LMC. And one of these two is a known Cepheid variable with a period of 39.345~days (OGLE ID LMC128.7 12809), which would require a mean absolute magnitude of $M_V=-5.5$ \citep[based upon the period/luminosity relationship for Cepheids, see e.g.,][]{Benedict2002}, placing it at the distance of the LMC.

For the SMC region, such an analysis is harder because its radial velocity is similar to the edge of the Milky Way distribution. But there are no stars in the SMC region which were identified as MC supergiants that have velocities higher than the expected range for the SMC. There are also no supergiant stars in the colour-colour diagram that are located in the region dominated by non-MC stars, as found in the LMC region.

Of interest are the 10 stars in the field region that passed the MC supergiant criteria, because these could be possible Magellanic Bridge or Stream members. Most are bluer stars (i.e., $\life>1$) as would be expected because there are few Ba-rich field stars. Being conservative, we classify six of these stars as false positives because they are found on the Galactic star sequences in the colour-colour diagram. These stars also happen to have proper motions barely inside the allowed region, i.e., they are on the edge of the red circle for proper motion selection (fourth row of Figure \ref{fig:mg_ba_vsini}).

Overall, these results suggest that there is a very low level of contamination from false positives. From the LMC there are less than one per cent (3/434) obvious false positives. For the field regions the contamination rate is much higher (6/10) but there were many more stars observed in the field regions than in the MC regions. False positives stars would likely be spatially random on the sky, so when observing a larger area, as we have done in the field regions, it is not unexpected that we would have more false positives.

\subsection{False negatives}\label{sec:false_negatives}
Along with the false positives, it is important to understand the rate of false negatives; are we rejecting MC members with \textit{The Cannon} label criteria? Or with the proper motion criteria? 

Although they are less certain than the radial velocities, the proper motions do provide a strong constraint. From the results using the entire SkyMapper sample of likely MC blue supergiants, 87 per cent of the $\sim3000$ stars were within 3.5~mas\,yr$^{-1}$ of the mean proper motion of the combined MC grouping.

There are eight stars that failed the proper motion criteria, but which passed the label criteria and that are in the radial velocity range of the LMC. Most of the other proper motion outliers are found in the the region of colour-colour space where the field dwarf and giant populations are found. All have relative proper motions larger than 6~mas\,yr$^{-1}$ from the rest of the LMC sample. This equates to a projected velocity on the plane of the sky of over $1000$~km\,s$^{-1}$ for a star with a relative proper motion $>5$~mas\,yr$^{-1}$, assuming it is at the distance of the LMC. Such speeds are occasionally seen in the Milky Way with hypervelocity stars escaping the Galaxy \citep[e.g.,][]{Guillochon2015}, but it seems unlikely we have discovered a population of hypervelocity MC stars, and instead these stars are considered to be Galactic stars coincident on the sky with the MC.

For the SMC, there are $15$ stars that were rejected because of their proper motions. All but two of these are found at such large relative proper motions that they cannot be considered associated with the SMC. One of the two stars that remain is most likely associated with the Galactic globular cluster NGC~362. NGC~362 does have a velocity similar to the MC \citep[$v_r=224$~km\,s$^{-1}$;][]{Harris1996} but with a proper motion slightly different to the SMC. There were $\sim20$ members of NGC~362 observed by GALAH, and this was the only one of these stars that passed the label criteria (in this case due to having a very low \mgfe).

Along with stars rejected because of their proper motions, there were stars rejected for their label values. In the LMC region there were 17 stars within the proper motion region, and that had radial velocities within the expected range for the LMC, but failed the label test. All but five of these failed the criteria that related to their \vsinidag.  Two of these five stars have very low \bafe\ but $\life\approx0.1$. Inspection of their other parameters suggests that they probably are legitimate members of the LMC: they are well above the \mgfe-\vsinidag\ line and, for the one that has SkyMapper colours, it is found at the very hot end of the supergiant sequence in $u-v$. It would appear that their \life\ is anomalously ``low'' (remembering that in all these OB supergiants \textit{The Cannon} has not detected a line but still has $\life\sim2$).

As discussed in Section \ref{sec:data_reduction}, one of the problems encountered by the reduction pipeline in measuring the radial velocity of the stars was that the very hottest stars have H-alpha emission and a strong He~I line at 6678~\AA. The radial velocities are determined using cross-correlation to 15 synthetic AMBRE spectra, and this was confusing this He line with the H-alpha line. This returns a velocity of $\sim5000$~km\,s$^{-1}$ for the red camera spectrum.  As such we have worked with the radial velocity from the blue camera of HERMES, which did not suffer from this problem. There were only two stars with blue camera radial velocity $v_r>80$~km\,s$^{-1}$ and a red camera velocity $\sim5000$~km\,s$^{-1}$ that were within the proper motion region that were not identified nor selected as supergiants: the two stars mentioned above with the low \life. There was another star that had an extremely large relative proper motion: $(\mu_l\cos b,\mu_b)=(29.9,10.9)$~mas\,yr$^{-1}$. Inspection of its spectra and colours shows it is actually a dwarf star that had a poor continuum normalization in the red spectra that caused problems for its radial velocity determination.

As with the false positives, it appears there are only a handful of false negatives based upon inspection of the colour-colour diagrams and the radial velocity distributions. Overall, it seems likely there are $<5$ false negatives of the several hundred stars identified as members in our sample.

\section{Stars of the Magellanic Bridge and Leading Arm}\label{sec:magellanic_bridge}

\begin{figure*}
    \includegraphics[width=\textwidth]{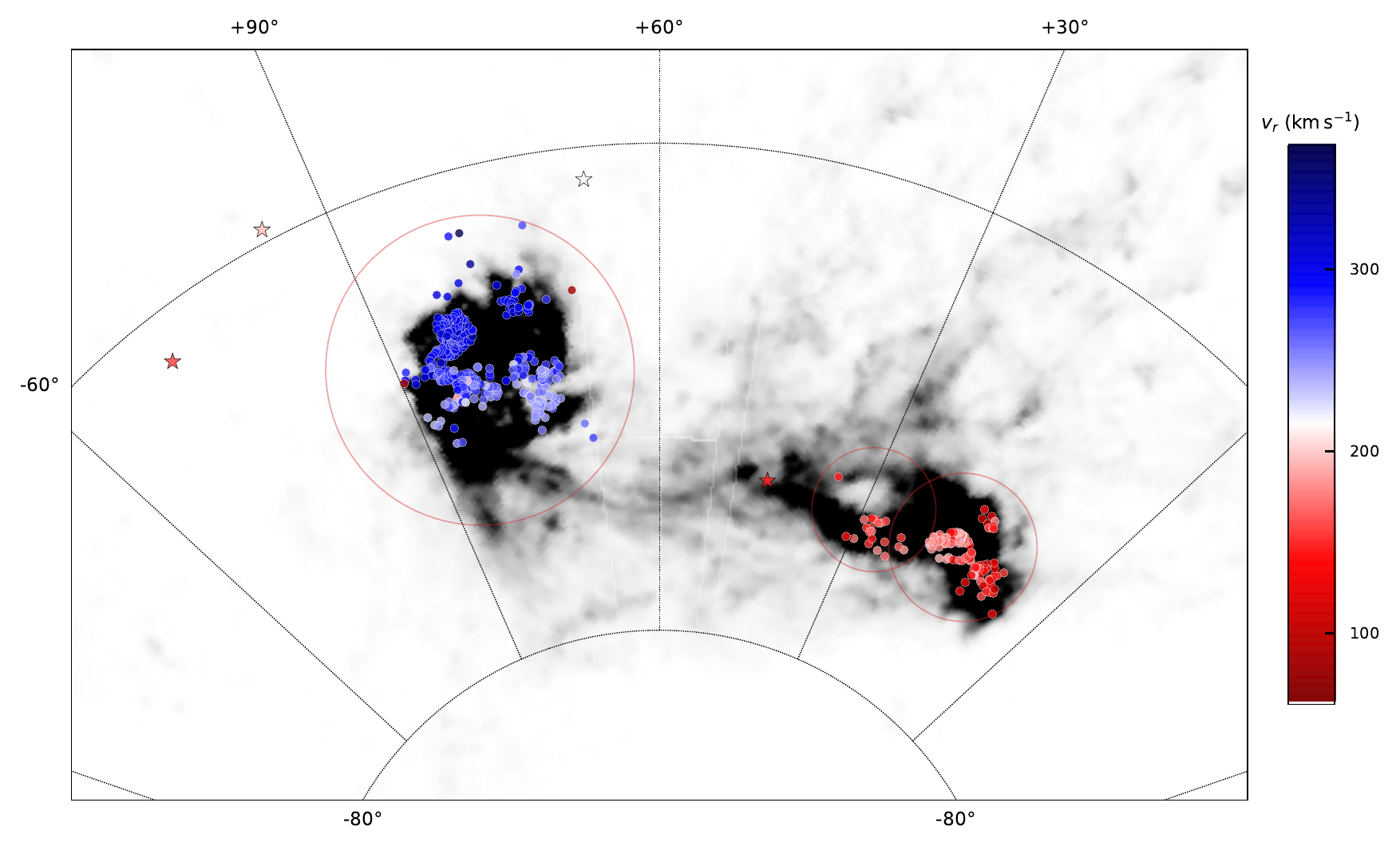}
    \caption{A map of HI emission integrated over a velocity range of $100<v_r<350$~km\,s$^{-1}$ from the Parkes Galactic All-Sky survey \citep[GASS;][]{McClure-Griffiths2009,Kalberla2010,Kalberla2015}. The MC are readily visible in the gas distribution, as well as the extended structure of the Magellanic bridge and streams. Also plotted are all the stars identified as MC supergiants coloured by their radial velocity. The four stars outside our nominal MC regions that have MC-like kinematics and label values are shown with star symbols.}
    \label{fig:galah_colour_sky}
\end{figure*}

Our study identified four stars that are located outside of the MC regions that met the criteria of having $v_r>80$~km\,s$^{-1}$, a small relative proper motion to the MC, \textit{Cannon} labels consistent with being a supergiant, and that were not as excluded as  Galactic stars due to their SkyMapper photometry. Figure \ref{fig:galah_colour_sky} shows the location of these stars on top of a map of HI gas column density from the Parkes Galactic All-Sky survey \citep[GASS;][]{McClure-Griffiths2009,Kalberla2010,Kalberla2015}. The symbols for each star are coloured by their radial velocity.

Of particular interest is the inter-MC star closest to the SMC. It has labels, photometry, and a spectrum consistent with being a young blue star, which would imply it is too young to have been stripped from the LMC or SMC. And this star is coincident with structured star formation in the region between the MC found by \citet{Mackey2017} in the form of strongly spatially clustered low-mass associations. One of the most striking features was a core-shell structure and this one star observed by GALAH is co-located with it. \citet{Mackey2017} found that this structure sat in a void in the surrounding HI gas, which could be interpreted as stellar winds and supernovae pushing away the gas. This particular star has a velocity of $145$~km\,s$^{-1}$, which is much slower than the speed of the gas with the maximum column density at its location, which is about 200~km\,s$^{-1}$. Another young star that has been observed spectroscopically, DI1388, is also coincident with the core-shell structure and was found by \citet{Hambly1994} to have a radial velocity of $v_r=150\pm30$~km\,s$^{-1}$.

What of the other three stars? There is evidence for an extended and lumpy stellar debris distribution around the MC, with over-densities of blue horizontal branch stars out to 30~deg from the LMC \citep{Belokurov2016}. The stars found here are all located on the far side of the LMC from the SMC, but have SMC-like radial velocities. Although the metallicities found by \textit{The Cannon} are not necessarily reliable, we have shown that \textit{The Cannon} has derived different metallicities for the two MC when considering just the red stars. These three stars are red and have low metallicities, with $\feh<-0.7$. \textit{The Cannon} found all the LMC stars to have $\feh>-0.5$, while the SMC was $-0.2>\feh>-0.9$. So this presents the intriguing possibility that these are cool supergiant stars stripped from the SMC, like those stars found by \citet{Carrera2017}. But they are quite distant from the SMC: the closest star is about 23~deg away on the sky, which would equate to a physical separation of at least 25~kpc, so the possibility is unlikely.

In this work, we have limited ourselves to stars within 15~deg of either MC. This has excluded the full extent of structures such as the Leading Arm, where young OB-type stars have been spectroscopically observed \citep{Casetti-Dinescu2012,Zhang2017}. Can we use our criteria to explore this region and identify these types of stars? As with the MC and Bridge, the Leading Arm has a high radial velocity relative to the Milky Way: $>130$~km\,s$^{-1}$ \citep{Zhang2017}. Any young star observed in the Leading Arm must be formed in situ: the distance from the LMC to the star would require it to have an ejection speed of $\sim10^4$~km\,s$^{-1}$. Taking the entire GALAH dataset, and applying our criteria, with the further constraint that stars must have $\life>0.9$ (i.e., the blue stars) and $v_r>150$~km\,s$^{-1}$, we find 28 stars scattered across the sky. None are particularly close to the structures explored by \citet{Zhang2017} and are likely fast moving OB stars in our Galaxy, ejected from their birth regions.

\section{Discussion}
In this work we have presented a search of the GALAH and \textit{TESS}-HERMES surveys for serendipitously observed MC stars. This work took advantage of not only the peculiar kinematics of MC stars with respect to the Milky Way, but the coherency of labels derived by \textit{The Cannon} from their spectra. \textit{The Cannon} is a data-driven method for measuring stellar labels from stellar spectra in the context of large spectroscopic surveys, which relies on a training set of spectra with known labels. The stars observed by GALAH and \textit{TESS}-HERMES in the MC are well outside of the training set, but \textit{The Cannon} is still able to place these stars in consistent (if not necessarily astrophysical) locations in the label space.

\textit{The Cannon} methods have been applied to spectra from GALAH \citep{Martell2017a}, \textit{TESS}-HERMES \citep{Sharma2017a}, K2-HERMES \citep{Wittenmyer2017}, APOGEE \citep{Ness2015,Casey2016}, LAMOST \citep{Ho2016a}, and RAVE \citep{Casey2016a}, but the emphasis in these works was on the accuracy and precision of the labels. This is the first work to explore in detail the usefulness of precise, but inaccurate, labels for stars that are far outside \textit{The Cannon} training set. \citet{Ness2015} do briefly discuss ``failures'' (as they referred to them) noting that their training set consisted almost entirely of giant stars. Their implementation of \textit{The Cannon} interpreted hot rotating dwarfs as very metal poor, cool giants.

We have also shown that \textit{The Cannon} has the potential to eventually derive meaningful information in the regime of luminous supergiants. There has been a lot of work showing that \textit{The Cannon}'s methodology works on main-sequence turnoff stars and red giant stars, but those are different kinds of stars. \textit{The Cannon} might have never worked for the kinds of stars discussed in this work if the underlying parameters were too sensitive to small changes, if they change in a non-quadratic way, or if there were additional covariances. The coherence of the parameters shows that there is a potential to eventually measure abundances and atmospheric parameters for these stars. The decision by the GALAH survey to not include colour cuts in its selection function was primarily driven by the desire to avoid making the selection function complicated. It was always known that this would result in the observation of stars that were too hot, or too cool for the abundance pipeline. But it was known that these spectra would be useful to someone or could be analyzed in the future with the extension of stellar models into these regimes. 

Astrophysically impossible \textit{Cannon} labels could be used to search for other classes of objects that lie outside the volume of parameter space occupied by the training set for GALAH, \textit{TESS}-HERMES and K2-HERMES. The problem of identifying extremely luminous stars in the MC is made a bit easier by their particular kinematics, especially their narrow range in proper motion. However, the foreground contamination was still significant in the field regions surrounding the MC, where dwarf stars managed to pass through our filters based on kinematics and \textit{Cannon} labels.

Future projects using similar methodology will be most approachable when they can also choose target samples with unusual properties relative to the overall sample. One promising group is white dwarfs, for which there is not yet a comprehensive Southern sky catalogue. In order to be observed by GALAH, they must be quite nearby, which will be easily verifiable with \textit{Gaia} DR2 data. We would also expect white dwarfs to be distinct in \textit{Cannon} label space because their spectra are quite unlike the majority of GALAH stars, and they can be identified quite clearly using the t-SNE dimensionality reduction technique on the spectra \citep{Traven2017}.

Large-scale observational projects continue to increase the volume and dimensionality of the data we have on Milky Way stars astrometrically, photometrically, spectroscopically, and asteroseismically. There is much to be learned by selecting populations of stars we believe to be well-understood in one of those spaces and tracking them through the other spaces to explore their full complexity. There is even more to be learned from following the natural distribution of the data, both in the correlations and dependencies between previously unconnected data sets and at the edges of the distribution, where outliers carry information about rare events in Galactic history.

\section*{Acknowledgements}
Based on data acquired through the Australian Astronomical Observatory, under programmes: A/2013B/13 (The GALAH pilot survey); A/2014A/25, A/2015A/19, A2017A/18 (The GALAH survey); A/2016B/10 (The \textit{TESS}-HERMES survey). We acknowledge the traditional owners of the land on which the AAT stands, the Gamilaraay people, and pay our respects to elders past and present.

The following software and programming languages made this research possible: \textsc{configure} \citep{Miszalski2006}; \textsc{iraf} \citep{Tody1986,Tody1993}; Python (versions 3.6); \textsc{astropy} \citep[version 3.0;][]{Robitaille2013,TheAstropyCollaboration2018}, a community-developed core Python package for Astronomy; \textsc{pandas} \citep[version 0.20.2;][]{McKinney2010}; \textsc{topcat} \citep[version 4.4;][]{Taylor2005}. This research made use of the cross-match service provided by CDS, Strasbourg.

DMN was supported by the Allan C. and Dorothy H. Davis Fellowship. SLM acknowledges support from the Australian Research Council through grant DE140100598. SB and KL acknowledge funds from the Alexander von Humboldt Foundation in the framework of the Sofja Kovalevskaja Award endowed by the Federal Ministry of Education and Research. LD gratefully acknowledges a scholarship from Zonta International District 24 and support from ARC grant DP160103747. KL acknowledges funds from the Swedish Research Council (Grant nr. 2015-00415\_3) and Marie Sklodowska Curie Actions (Cofund Project INCA 600398). TZ acknowledge the financial  support  from  the  Slovenian  Research  Agency  (research core funding No. P1-0188). LD, KF and Y-ST are grateful for support from Australian Research Council grant DP160103747. Parts of this research were conducted by the Australian Research Council Centre of Excellence for All Sky Astrophysics in 3 Dimensions (ASTRO 3D), through project number CE170100013. 

The national facility capability for SkyMapper has been funded through ARC LIEF grant LE130100104 from the Australian Research Council, awarded to the University of Sydney, the Australian National University, Swinburne University of Technology, the University of Queensland, the University of Western Australia, the University of Melbourne, Curtin University of Technology, Monash University and the Australian Astronomical Observatory. SkyMapper is owned and operated by The Australian National University's Research School of Astronomy and Astrophysics. The survey data were processed and provided by the SkyMapper Team at ANU. The SkyMapper node of the All-Sky Virtual Observatory (ASVO) is hosted at the National Computational Infrastructure (NCI). Development and support the SkyMapper node of the ASVO has been funded in part by Astronomy Australia Limited (AAL) and the Australian Government through the Commonwealth's Education Investment Fund (EIF) and National Collaborative Research Infrastructure Strategy (NCRIS), particularly the National eResearch Collaboration Tools and Resources (NeCTAR) and the Australian National Data Service Projects (ANDS).











\bsp	
\label{lastpage}
\end{document}